\documentclass[12pt]{article}
\pdfoutput=1
\usepackage{epsfig}
\usepackage{amsfonts}
\usepackage{amssymb}
\usepackage{xcolor}

\begin{document}
\newcommand{\nc}{\newcommand}
\nc{\beq}{\begin{equation}} \nc{\eeq}{\end{equation}}
\nc{\beqa}{\begin{eqnarray}} \nc{\eeqa}{\end{eqnarray}}
\nc{\R}{{\cal R}}
\nc{\A}{{\cal A}}
\nc{\K}{{\cal K}}
\nc{\B}{{\cal B}}
\begin{center}

{\bf \Large  Kinematically Dependent Renormalization} \vspace{1.0cm}

{\bf \large D. I. Kazakov$^{1,2}$} \vspace{0.5cm}

{\it
$^1$Bogoliubov Laboratory of Theoretical Physics, Joint
Institute for Nuclear Research, Dubna, Russia.\\
$^2$Moscow Institute of Physics and Technology, Dolgoprudny, Russia\\}
\vspace{0.5cm}

\abstract{We suggest a renewed view on non-renormalizable interactions treated perturbatively
within  a kinematically dependent renormalization procedure. It is based on the usual  BPHZ $\R$-operation which is equally applicable  to any local QFT independently whether it is renormalizable or not. The key point is that the renormalization constant becomes the function of kinematical variables acting as an operator on the amplitude. The procedure is demonstrated by the example of D=8 supersymmetric gauge theory considered within the spinor helicity formalism.}
\end{center}


\section{Introduction}
As is well known, non-renormalizable interactions suffer from UV divergences and cannot be treated in a usual renormalization fashion due to the fact that in each order of perturbation theory one has new divergent structures which do not repeat the original Lagrangian. These new structures contain higher powers of momenta in momentum space or higher derivatives in the coordinate space to compensate the negative dimension of the coupling. Consequently, to subtract the divergence and fix the subtraction arbitrariness, one has a new condition at each order of PT for each new operator. 
This leads to infinite arbitrariness in the subtraction procedure which is not acceptable. Unless all these divergences are related to one another. Just this property we suggest to explore in what follows.

Any local  QFT has the property that in higher orders of PT after subtraction of divergent subgraphs the remaining UV divergences are local functions in the coordinate space or at maximum are polynomials of external momenta in momentum space. This follows from the rigorous proof of the Bogoliubov-Parasiuk-Hepp-Zimmermann $\R$-operation~\cite{BPHZ}. This statement is equally valid in non-renormalizable theories as well. The immediate consequence of this statement is that the divergent terms proportional to $\log^k s/\mu$, where $s$ is the kinematical variable and $\mu$ is the renormalization scale, which inevitably appear in the calculation of  higher order diagrams, must cancel as a result of subtraction of divergent subgraphs. Indeed, this  happens for each individual diagram and for the whole series in a given order of PT.  

However, this cancellation is possible only if the higher order divergences, which give rise to such non-local terms, are not independent but are related to the lower order ones. These relations indeed exist and are governed by the renormalization group equations in  renormalizable theories but are equally valid in non-renormalizable ones, though take a more sophisticated form. When using dimensional regularization these relations are nothing more than the  't Hoofts pole equations~\cite{tHooft}, which relate the higher order poles in $\epsilon$ with the lower order ones. In a recent set of papers~\cite{we1,we2} we demonstrated how these generalized pole equations could be written in the case of D=6, 8 and 10 dimensional super Yang-Mills theories, which are non-renormalizable by power counting. 

The aim of this letter is to suggest the reasoning that at least in some cases the non-renormalizable interactions can be treated perturbatively exactly in the same way as the renormalizable ones. The  only difference is that the renormalization constant $Z$ becomes the function of kinematic variables  and acts on the amplitude not as a simple multiplication but as the operator in momentum space.
This {\it kinematically dependent renormalization} is performed exactly in the same way as the usual one in a sense that all the UV divergences in the amplitudes are removed  by  applying  the renormalization constant and replacing the bare coupling with the renormalized one.  Transition from one subtraction scheme to another is governed by the finite renormalization constructed along the same lines. Below we demonstrate this procedure by the example of D=8 supersymmetric gauge theory considered within the spinor helicity formalism.

\section{D=8 SYM theory in spinor helicity formalism}

As a playground for our analysis we choose the planar scattering amplitudes in D=8 super Yang-Mills theory considered within the spinor helicity formalism~\cite{Reviews_Ampl_General}. One does not need to know the details of this formalism, and the choice of a given theory is not essential for our analysis.  It is just the simplicity of the on-shell scattering amplitudes in the spinor helicity formalism that allows one to make calculations up to several loops and explicitly  trace all steps of the proposed procedure. The coupling $g^2$ in this case has the mass dimension equal to -4, so the theory is non-renormalizable by power counting.

\begin{figure}[htb]
\includegraphics[scale=0.35]{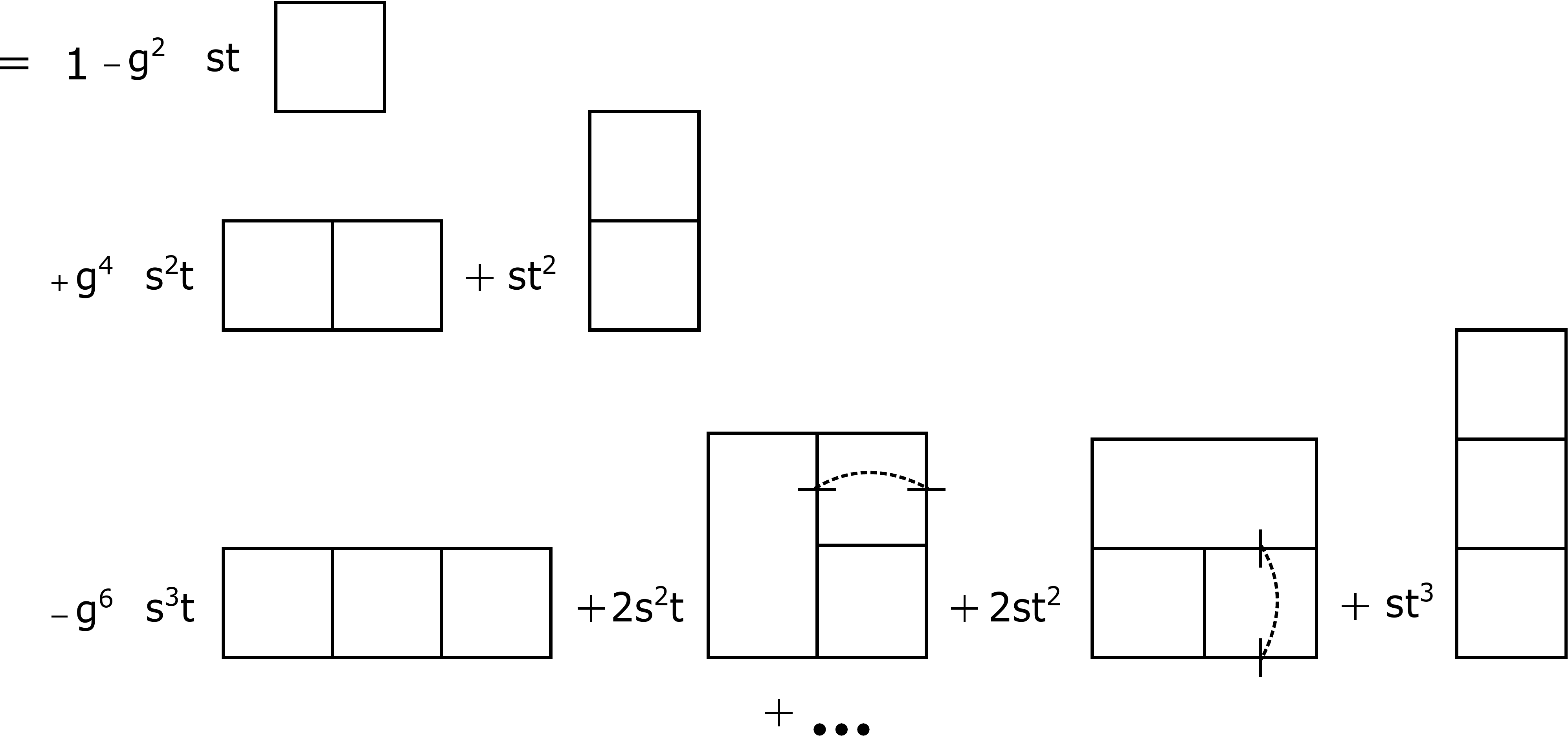}
\caption{The universal expansion for the four-point scattering amplitude in SYM theories in terms of master integrals.  The connected strokes on the lines mean  the square of the flowing momentum.
}\label{expan}
\end{figure}

Let us remind first how the PT series for the four-point scattering amplitude looks like in this case. All simple loops cancel and one has mainly the box type diagrams (see Fig.\ref{expan}). Since we consider the on-shell amplitudes, all $p_i^2=0$ and one has only dependence on the Mandelstam variables
$s$ and $t$  in each  channel. Using dimensional regularization, one can calculate the divergent parts of the diagrams order by order in PT. Explicit evaluation gives
\beq
Box=\frac{1}{3!\epsilon}, \ \ \  \ DoubleBox=-\frac{1}{3!4!}(\frac{s}{\epsilon^2}+\frac{27}{4}\frac{s}{\epsilon}+\frac 16\frac{t}{\epsilon}), \ \ \ etc.
\eeq
\beqa
\bar{\A}_4= \frac{\mathcal{A}_4}{\mathcal{A}_4^{(0)}}=1+\sum\limits_L M^{(L)}_4(s,t)= 
\eeqa \vspace{-0.7cm}

\noindent Substituting these integrals into the normalized amplitude $\bar{\A}_4$, one gets for the singular part:
\beq
\bar{\A}_4= 1-\frac{g^2 st}{3!\epsilon}-\frac{g^4 st}{3!4!}\left(\frac{s^2+t^2}{\epsilon^2}+\frac{27/4 s^2+1/3 st +27/4 t^2}{\epsilon}\right) + ...  \label{gamma}
\eeq

Note that the coefficients of the higher order poles are not arbitrary but totally governed by the lower
order ones due to the above mentioned  cancellation of nonlocal structures. This statement can be quantified in the form of the pole recurrence relations which allow one to evaluate the higher order poles algebraically. For the leading divergences in D=8 SYM theory these relations were derived in~\cite{we1} and take the form
\beqa
&&nS_n(s,t)=-2 s^2 \int_0^1 dx \int_0^x dy\  y(1-x) \ (S_{n-1}(s,t')+T_{n-1}(s,t'))|_{t'=tx+yu}\label{eq8}\nonumber \\ &+&
s^4 \int_0^1\! dx \ x^2(1-x)^2 \sum_{k=1}^{n-2}  \sum_{p=0}^{2k-2} \frac{1}{p!(p+2)!} \
 \frac{d^p}{dt'^p}(S_{k}(s,t')+T_{k}(s,t')) \times \nonumber \\
&&\hspace{2cm}\times  \frac{d^p}{dt'^p}(S_{n-1-k}(s,t')+T_{n-1-k}(s,t'))|_{t'=-sx} \ (tsx(1-x))^p,
\label{recur}
\eeqa
where $t'= t x+u y$, $u=-t-s$, and $S_1= \frac{1}{12},\ T_1=\frac{1}{12}$.
Here we denote by  $S_n(s,t)$ and  $T_n(s,t)$  the sum of all contributions  in the  $n$-th order of PT in $s$ and  $t$ channels, respectively.
Thus, substituting the one loop values  $S_1$ and $T_1$ into eq.(\ref{recur}) one gets the two loop higher poles
$$S_2=-\frac{s^2}{3!4!}, \ \  T_2=-\frac{t^2}{3!4!}, $$
in accordance with eq.(\ref{gamma}).

The complexity of eq.(\ref{recur}), compared to the usual pole relations in renormalizable theories, originates from the fact that the pole terms are no more the constants but are the polynomials of external momenta. When calculating the  subdivergences of multiloop diagrams, these momenta
being external for the subgraph become the internal momenta for the whole graph and have to be integrated out. The integrals entering into eq.(\ref{recur}) have the meaning of integration over the  Feynman parameters of subgraphs which appear after shrinking the subgraph to a point in due course of the $\R$-operation. This closed equation summarizes the diagram by diagram calculation with the help of the well defined $\R$-operation.

One can continue this procedure for the subleading divergences where now both the one loop and the two loop genuine contributions to the  simple pole have to be taken as initial conditions~\cite{we2}. The relations are too cumbersome to write them down here but work perfectly well as the ones for the leading poles. The essence of the statement is that knowing the lowest diagrams, one can evaluate the leading, subleading, etc divergences pure algebraically since they are not independent but follow from the $\R$-operation.

\section{Kinematically dependent renormalization procedure}
We now describe the renormalization procedure and illustrate it by the example of the two loop divergences in D=8 SYM theory. Formally, it looks {\it precisely} like a familiar renormalization procedure in any renormalizable theory. Namely, to get a finite amplitude, one has to multiply
the bare amplitude by the proper renormalization constant and replace the bare coupling 
with the renormalized one, according to the following formulas:
\beqa
\bar{\A}_4& = &Z_4(g^2) \bar{\A}_4^{bare}|_{g^2_{bare}->g^2Z_4}, \label{mult}\\
g^2_{bare}&=&\mu^\epsilon Z_4(g^2)g^2. \label{coupling}
\eeqa
Remind also that the renormalization constant $Z_4$ can be calculated diagrammatically with the help
of the following standard operation~\cite{Rop}:
\beq
Z  = 1-\sum_i\K\R' G_i, \label{ZZ}
\eeq
where  $\R'$ is the incomplete $\R$-operation,  which subtracts only  the subdivergences of the graph $G_i$, and $\K$ is an operator that singles out the singular part of the graph (for the minimal subtraction scheme  the operator $\K$ singles out the  $1/\epsilon^n$ terms). The
$\K\R' G_i$  is the counter term corresponding to the  graph $G_i$.
Each counter term contains only the superficial divergence and is  {\it local} in the coordinate space (in our case it must be a polynomial of external momenta).

To demonstrate  how the renormalization procedure works and to see the essential difference between the non-renormalizable and the renormalizable cases,
we apply the  renormalization procedure (\ref{mult},\ref{coupling}) to the amplitude 
(\ref{gamma}) order by order in PT.  

\underline{1loop order}: The coupling is not changed in this order  $g^2_{bare}=\mu^\epsilon g^2$
and the renormalization constant is chosen in the form $Z_4=1+\frac{g^2 st}{3!\epsilon}$. This leads to a finite answer. Notice that the renormalization constant is not really a constant but depends on the kinematic factors $s$ and $t$! 

\underline{2 loop order}: The coupling is changed now according to (\ref{coupling}), namely,
\beq
g^2_{bare}=\mu^\epsilon g^2(1+\frac{g^2 st}{3!\epsilon}) \label{1lcoup}\eeq
 and the renormalization constant is taken in the form
\beq
Z_4=1+\frac{g^2 st}{3!\epsilon}+\frac{g^4 st}{3!4!}\left(\frac{A_2s^2+B_2st+A_2t^2}{\epsilon^2}+
\frac{A_1s^2+B_1st+A_1t^2}{\epsilon}\right),\label{2lz}
\eeq
where the coefficients $A_i$ and $B_i$  have to be chosen in a way to cancel all divergences both local and nonlocal ones.  

Now comes the key point of our analysis. When substituting eqs.(\ref{1lcoup},\ref{2lz}) into eq.(\ref{mult}), one can notice that replacement of $g^2_{bare}$ by expression (\ref{1lcoup}) in the one loop term ($\sim g^2$) and multiplication of one loop contributions from the renormalization constant $Z_4$  and from the amplitude $\bar{\A}_4$ have the effect of subtraction of subdivergences in the two loop graph. This is exactly what guarantees the locality of the counter terms within the $\R$-operation. However, contrary to the renormalizable case, here the renormalization constant contains the kinematic factors, the powers of momenta, which are external momenta for the subgraph but become  internal ones for the remaining diagram. This means that they have to be inserted inside the remaining diagram and integrated out. To clarify this point, we consider the corresponding term which appears when multiplying the one loop Z factor on the one loop amplitude. The $s$ and $t$ factors from the Z factor have to be inserted inside the box diagram
\begin{figure}[h]
\begin{center}
\includegraphics[scale=0.50]{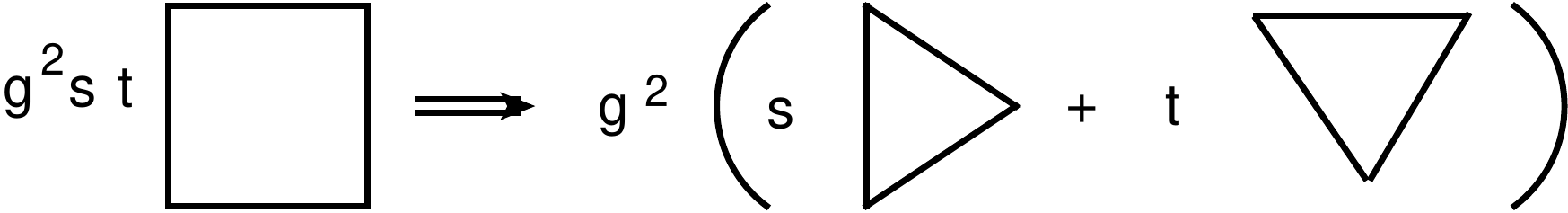}
\end{center}
\end{figure}

This means that the simple {\it multiplication} procedure has to be modified: the Z factor becomes the {\it operator} acting on the diagram which inserts the powers of momenta inside the diagram. This looks a bit artificial but exactly reproduces the $\R$-operation for the two loop diagram shown below
in Fig.\ref{2loop}.
\begin{figure}[h]
\begin{center}
\includegraphics[scale=0.45]{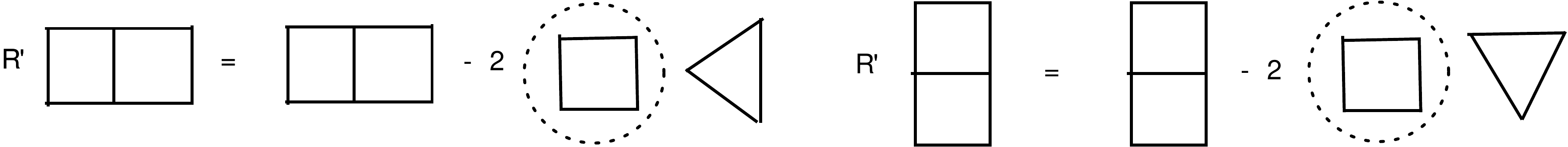}
\caption{$\R'$-operation for the two loop diagrams}\label{2loop}
\end{center}
\end{figure}

Let us see now how this operation works for the four-point amplitude in the two loop order. Inserting eqs.(\ref{2lz},\ref{1lcoup}) into eq.(\ref{mult}) and having in mind that 
\beq
sTriangle=-\frac{s}{4!\epsilon}(1+\frac{19}{6}\epsilon), \ \  tTriangle=-\frac{t}{4!\epsilon}(1+\frac{19}{6}\epsilon),
\eeq
one gets
\beqa
\bar{\A}_4& = &Z_4(g^2) \bar{\A}_4^{bare}|_{g^2_{bare}->g^2Z}\nonumber \\
&=& 1-\frac{g^2 \mu^\epsilon st}{3!\epsilon}+\frac{g^2 st}{3!\epsilon}-\frac{g^4 \mu^{2\epsilon} st}{3!4!}\left(\frac{s^2+t^2}{\epsilon^2}+\frac{27/4 s^2+1/3 st +27/4 t^2}{\epsilon}\right)  \\
&+& 2\frac{g^4 st}{3!\epsilon} \mu^{\epsilon}\frac{s^2+t^2}{4!\epsilon}(1+\frac{19}{6}\epsilon)+
\frac{g^4 st}{3!4!}\left(\frac{A_2s^2+B_2st+A_2t^2}{\epsilon^2}+
\frac{A_1s^2+B_1st+A_1t^2}{\epsilon}\right).\nonumber
\eeqa
One can see that the one loop divergences ($\sim g^2$) cancel and the cancellation of the two loop ones requires
\beqa
\frac{1}{\epsilon^2}: && -\frac{s^2+t^2}{3!4!}st +2 \frac{s^2+t^2}{3!4!}st + \frac{A_2s^2+B_2 st +A_2 t^2}{3!4!}st=0,\nonumber \\
\frac{\log\mu}{\epsilon}: &&  -2\frac{s^2+t^2}{3!4!}st +2 \frac{s^2+t^2}{3!4!}st =0,\nonumber \\
\frac{1}{\epsilon}: &&-\frac{st}{3!4!}(\frac{27}{4} s^2+\frac 13 st +\frac{27}{4} t^2)+2\frac{st}{3!4!}(s^2+t^2)\frac{19}{6}
+\frac{st}{3!4!}(A_1s^2+B_1st+A_1t^2)=0. \nonumber
\eeqa
One deduces that $A_2=-1,B_2=0, A_1=\frac{5}{12}, B_1=\frac 13$, so that the renormalization constant $Z_4$ takes the form
\beq
Z_4=1+\frac{g^2 st}{3!\epsilon}+\frac{g^4 st}{3!4!}\left(-\frac{s^2+t^2}{\epsilon^2}+
\frac{5/12 s^2+1/3st+5/12 t^2}{\epsilon}\right),\label{2lzp}
\eeq
which exactly corresponds to the one obtained using eq.(\ref{ZZ}). This expression now has to be substituted into eq.(\ref{coupling}) to obtain the renormalized  coupling. Note that it also depends on kinematics. 

In fact, this means that we build this way an induced higher derivative theory where the higher terms appear order by order of PT with fixed coefficients. For instance, the one loop term
$g^2 s t/\epsilon$ generates the gauge invariant counter term 
$$ \frac{g^2}{\epsilon}D_\rho D_\lambda F_{\mu\nu}D_\rho D_\lambda F_{\mu\nu},$$
that contains higher derivatives as well as new vertices with extra gauge fields, etc.

The whole construction obviously depends on the subtraction procedure. We used the minimal subtraction scheme so far but can equally use another one. For instance, if one introduces an arbitrary subtraction constant  $c_1$ in the one loop box diagram, it will enter into all the subleading divergences. Moreover, the part of subleading divergences proportional to $c_1$ is given by the derivative of the leading ones with respect to the coupling $g^2$. This means that  it is generated by the shift of the coupling $g^2 \to g^2(1+c_1\epsilon )$ in the leading term. This fact was first noticed in~\cite{we2} and  elaborated in~\cite{we5}. In the context of the present discussion the transition to a non-minimal scheme is equivalent to the multiplication of the amplitude by the finite renormalization constant 
 \beq z = 1+g^2 s t c_1 \label{fin}
 \eeq
 and the corresponding finite change of the coupling $g$.
 This looks similar to the renormalizable case though the meaning is different. Again, it is not simply the multiplication  but  the action of the operator which is also kinematically dependent.  Therefore, it is not a simple change of a single coupling but  of the whole infinite series of higher derivative terms.
 
 Similarly, the subtraction arbitrariness of the double box influences the subsubleading divergences and results in higher order terms in eq.(\ref{fin}) like in eq.(\ref{2lzp}),  just as in renormalizable theories~\cite{we5}.  
 Therefore, the whole arbitrariness is accumulated in one renormalization constant evaluated order by order in PT, which acts as the operator and generates an infinite series of terms.
 
 It is instructive to compare this point with the situation in the renormalizable theory. In the latter case, one can introduce arbitrary subtraction constants at each loop and as a result have an infinite number of them.  However, all of them are related to the normalization of a single operator term and eventually are absorbed into a single renormalization constant $z$. Here we also have an infinite number of subtraction constants. Contrary to the renormalizable case, they correspond to different operator terms. Still, even in this case, one can absorb them into a single operator $z$ acting on the amplitude.

\section{Discussion}

Let us try to summarize what we have got. 
\begin{itemize}
\item  First of all, the $\R$-operation works pretty well and allows one to get a finite answer.
\item Second, it can be equivalently formulated  via eqs.(\ref{mult},\ref{coupling}); however, it is not a simple multiplication now but the action of the operator $Z$ which depends on the kinematics. 
We have no rigorous proof of it, but fortunately it is not needed. The proof of the $\R$-operation is sufficient. Everything works diagram by diagram according to  the standard procedure. However, when one tries to formulate the general algorithm how the operator $Z$ acts on a given diagram, one meets with difficulties. The problem is manifested in eq.(\ref{recur}) which looks complicated enough though it  is only the leading  divergence.
\item Third, the scheme dependence is accomulated in  a single renormalization operator that 
depends on  kinematics.

\item And the last, and this is our key statement, {\it one can make sense of a non-renormalizable theory renormalizing a single coupling with the help of  the renormalization constant that depends on kinematics}. 
As a result, one can construct the higher derivative theory that gives the finite scattering amplitudes with a single arbitrary coupling $g$ defined in PT within the given renormalization scheme. Transition to another scheme is performed by  the action on the amplitude of a finite renormalization operator $z$ that depends on kinematics.

Does this procedure really fix the theory or is it merely an illusion remains unclear to us.

\item Assuming that one accepts these arguments, there is still a problem that at each order of PT the amplitude increases with energy, thus violating the unitarity.  However, apparently, this problem has to be addressed after summation  of the whole PT series. While each term of PT behaves badly, the whole sum might behave differently. We analyzed the behaviour of the amplitudes in D=6, 8 and 10 SYM theories in the leading order using the generalized RG equations in~\cite{we3,we4}. Each case is really different and one should consider this question  separately.   Some theories might be acceptable, some not. However, all this makes sense only if the key statement is correct.
\end{itemize}

\section*{Acknowlegments}
The author is grateful to colleagues from BLTP for useful discussions. This work was supported by the Russian Science Foundation grant \# 16-12-10306.

\end{document}